\documentclass[11pt]{article}

\usepackage[preprint]{acl}

\usepackage{times}
\usepackage{latexsym}
\usepackage[T1]{fontenc}
\usepackage[utf8]{inputenc}
\usepackage{microtype}
\usepackage{graphicx}
\usepackage{booktabs}
\usepackage{enumitem}
\usepackage{makecell}
\usepackage{xurl}

\hypersetup{%
  pdftitle={What Does It Cost to Simulate a Quantum Sentence Classifier? An Energy and Compute Perspective on Near-Term QNLP},
  pdfauthor={Sandipan Ganguly, Kishlay Kashyap}
}

\title{What Does It Cost to Simulate a Quantum Sentence Classifier?\\ An Energy and Compute Perspective on Near-Term QNLP}

\author{%
  \href{https://orcid.org/0009-0000-7502-9121}{Kishlay Kashyap} \\
  Heritage Institute of Technology \\
  \texttt{kishlaykashyap77@gmail.com}
  \And{}
  \href{https://orcid.org/0009-0005-3181-778X}{Sandipan Ganguly} \\
  Heritage Institute of Technology \\
  \texttt{ganguly.sandipan@gmail.com}
}

\begin{document}
\maketitle

\begin{abstract}
Near-term quantum natural language processing (QNLP) experiments often run on classical simulators, so simulator cost is part of the field's practical compute burden, yet accuracy tables do not show it. We measure that cost for a variational quantum classifier (VQC) on binary SST-2 sentiment classification, using PennyLane's state-vector simulator over a controlled grid of 27 configurations: three balanced training-set sizes ($N\in\{200,500,1000\}$), three qubit counts (4, 6, 8), and three circuit depths. Each VQC is compared with logistic regression on the same PCA-reduced input; full TF-IDF logistic regression gives an uncompressed reference. The VQC beats its matched baseline in 6 of 27 single-seed comparisons. After reruns at two further seeds, only 1 of these 6 keeps a positive mean advantage larger than its paired seed-to-seed variability, and paired tests on the fixed validation set do not establish it. VQC training is 886--21,127$\times$ slower in measured wall-clock time than the matched classical fit (median 3,158$\times$); going from 4 to 8 qubits roughly doubles simulator time, and within the tested range per-step cost is well approximated by a linear function of the parameter count. CodeCarbon energy and CO$_2$ estimates are secondary: they imply an almost constant power of about 41~W, so they add little beyond runtime, and we do not build an energy ratio from them. The study is narrow (one dataset, representation, ansatz, simulator, and CPU environment) and is a reproducible feasibility measurement, not a general verdict on QNLP\@.
\end{abstract}

\section{Introduction}
\label{sec:intro}

Quantum natural language processing (QNLP) asks whether quantum computational structure can supply useful inductive biases for language. It inherits two broad strategies from quantum machine learning \citep{biamonte2017qml,havlicek2019supervised}. Compositional approaches build a sentence's grammatical structure into a quantum circuit, following the categorical compositional distributional (DisCoCat) program \citep{coecke-etal-2010-mathematical} and its software realization in \texttt{lambeq} \citep{kartsaklis2021lambeq}; such models have been executed on quantum hardware for small sentence-classification tasks \citep{lorenz2023qnlp}. Embedding-style approaches instead encode a fixed-dimensional classical feature vector into a parameterized circuit and train it variationally, as in circuit-centric quantum classifiers \citep{schuld2020circuit}; they are closely related to quantum kernel methods \citep{havlicek2019supervised,schuld2021kernel}. Current processors are noisy, small, and of restricted access \citep{preskill2018nisq}, so much QNLP experimentation is carried out on classical simulators, with PennyLane one widely used framework \citep{bergholm2018pennylane}.

Simulation is itself a computational workload. A state-vector simulator stores $2^{n}$ amplitudes for $n$ qubits, so memory and arithmetic grow steeply with circuit width, and variational training multiplies the cost of one circuit evaluation by the number of optimizer steps. A result obtained ``on 6 qubits'' therefore hides the classical compute, and ultimately the electricity, that produced it. \citet{strubell-etal-2019-energy} estimated the energy and carbon cost of training NLP models. Later work proposed standardized reporting practices \citep{schwartz2020green,henderson2020systematic} and argued for reporting efficiency as an evaluation criterion. More recent is similar energy and carbon accounting for quantum-circuit simulation \citep{li2023carbon,tripathi2025carbon}. This is relevant in the interpretation of results since quantum machine learning benchmarks can vary significantly based on the experimental setup \citep{bowles2024better}.

This paper brings these threads together in one deliberately simple, fully logged experiment. For a VQC on binary SST-2 sentiment classification we (i) measure simulator cost under one timing definition applied to both quantum and classical arms, (ii) compare against a classical baseline matched on input dimensionality rather than given a different representation, and (iii) check whether apparent accuracy wins survive reruns at other seeds and paired statistical tests. We vary the training-set size $N\in\{200,500,1000\}$, the qubit count (4, 6, 8), and the circuit depth (1--3 layers), giving 27 configurations, and ask:

\begin{itemize}[leftmargin=1.5em,itemsep=1pt,topsep=2pt]
  \item \textbf{RQ1 (cost).} How much measured wall-clock time, and how much software-estimated energy, does simulating VQC training consume relative to a classical classifier on \emph{the same} compressed representation, and how does the cost scale with qubit count, depth, and training-set size?
  \item \textbf{RQ2 (accuracy).} Does the VQC beat its dimensionality-matched classical baseline under a fixed optimization protocol?
  \item \textbf{RQ3 (robustness).} Do single-seed accuracy comparisons survive targeted multi-seed reruns and paired tests on the fixed validation set?
\end{itemize}

Wall-clock time is the primary cost metric. CodeCarbon \citep{courty2024codecarbon} supplies secondary energy and CO$_2$ estimates, which we do not treat as precise electricity measurements. In brief: matched VQC/classical time ratios are 886--21,127$\times$ (median 3,158$\times$); the VQC leads in 6 of 27 single-seed comparisons, one of which keeps a positive mean advantage larger than its seed-to-seed variability after two further seeds, and none is statistically established.

The contribution is methodological rather than algorithmic:
\begin{itemize}[leftmargin=1.5em,itemsep=1pt,topsep=2pt]
  \item a reproducible matched-dimensionality cost measurement for a simulator-based QNLP classifier, with every quoted number recomputed from the exported artifacts by a verification script;
  \item a decomposition of simulator time by qubit count, depth, and optimizer steps, showing that at this width the 4$\rightarrow$8 qubit growth is about $2\times$, and that within the tested range per-step cost is well approximated by a linear function of the parameter count rather than of the $2^{n}$ state dimension;
  \item a targeted multi-seed protocol for apparent accuracy advantages, with explicit regression-to-the-mean caveats; and
  \item diagnostics for near-degenerate predictions, and evidence that software energy estimates carry essentially the same information as wall-clock time in this environment.
\end{itemize}

\section{Related Work}
\label{sec:related}

\paragraph{QNLP approaches.}
In DisCoCat, a syntactic parse specifies how distributional word representations are combined. The resulting tensor network can then be compiled into a quantum circuit \citep{coecke-etal-2010-mathematical}. \texttt{lambeq} implements this workflow as an end-to-end toolkit \citep{kartsaklis2021lambeq}, and small models of this type have been run on quantum hardware \citep{lorenz2023qnlp}. Our experiment uses a different formulation: the circuit is a parameterized function of a fixed-dimensional classical input vector \citep{schuld2020circuit}. Many classifiers in this family also admit a kernel interpretation \citep{schuld2021kernel}, so we draw no conclusions about grammar-aware QNLP architectures.

\paragraph{Trainability and benchmarking.}
Sufficiently expressive random circuits can exhibit barren plateaus, in which gradients vanish exponentially with system size \citep{mcclean2018barren}; we do not measure them and report optimization behavior as an empirical property of the tested stack. \citet{bowles2024better} benchmarked 12 quantum machine-learning models on 160 datasets, found that out-of-the-box classical models outperform the quantum classifiers, and stressed the large effect of experimental design on such comparisons. Our matched-dimensionality baseline, fixed protocol, and multi-seed reruns are small-scale responses in the same spirit.

\paragraph{Energy and carbon reporting.}
Energy reporting in NLP predates the present study. \citet{strubell-etal-2019-energy} estimated the energy use and carbon emissions of model training, and \citet{schwartz2020green} argued that accuracy should not be the only evaluation criterion. \citet{henderson2020systematic} followed with a tracking framework and a standardized reporting scheme for machine-learning experiments. Work on quantum simulation is less extensive. \citet{li2023carbon} proposed a preliminary model for the CO$_2$e of circuit simulation; under that model, a 43-qubit simulation could emit 48 times as much CO$_2$e as training a transformer. \citet{tripathi2025carbon} examined how the choice of ansatz changes this resource cost. In our experiment, wall-clock time is the main cost measure. We report CodeCarbon values only as software estimates and state the assumed grid intensity (India) explicitly.

\section{Task and Data}
\label{sec:data}

We use SST-2, the binary-sentiment split of the Stanford Sentiment Treebank \citep{socher-etal-2013-recursive}, distributed through GLUE \citep{wang-etal-2018-glue} and loaded from the Hugging Face \texttt{stanfordnlp/sst2} dataset.\footnote{\url{https://huggingface.co/datasets/stanfordnlp/sst2}} The training split contains 67,349 labeled examples (phrases and sentences; 29,780 negative and 37,569 positive); the fixed validation split contains 872 sentences (428 negative and 444 positive). The hidden-label test split is not used.

From the training split we construct deterministic, exactly balanced subsets of $N\in\{200,500,1000\}$ examples using the main seed 42, so class prior is not a confound within training. The same 872 validation examples evaluate every experiment, which makes accuracies directly comparable across qubit count, depth, and training-set size. Because the validation split is mildly imbalanced, the majority-class baseline is $444/872=0.5092$; we show it alongside the 0.5 chance reference.

\section{Experimental Design}
\label{sec:method}

\subsection{Feature Pipeline}
\label{sec:pipeline}
For each $N$, TF-IDF with unigrams and bigrams, sublinear term frequency, and a 3,000-feature vocabulary cap is fit on the training subset only and applied to validation, so no validation information enters the vocabulary or term weights. PCA is then fit on the training TF-IDF matrix and reduces the representation to 4, 6, or 8 components, matching the qubit count. From this single PCA output two representations are derived, each fit on training data only: a \emph{standardized} version for the classical arm and a \emph{linearly rescaled} version mapped to $[0,\pi]$ for the VQC arm, as \texttt{AngleEmbedding} expects rotation angles; validation values are clipped to the same interval. The two matched arms therefore consume the same reduced representation and differ only in their classifier.

At $N=1000$, the 4-, 6-, and 8-component PCA representations retain approximately 3.4\%, 4.7\%, and 5.9\% of total TF-IDF variance. This is an intentionally severe bottleneck, chosen because near-term qubit budgets force this kind of compression; since PCA is unsupervised, explained variance is not a measure of retained sentiment-predictive information, and the full-TF-IDF baseline is reported to show what the compression costs.

\subsection{Classical Baselines}
\label{sec:baselines}
\emph{Full TF-IDF logistic regression} is the uncompressed reference and uses all 3,000 features. The \emph{PCA-matched baseline} uses the same 4-, 6-, or 8-dimensional standardized representation given to the VQC\@. Both use scikit-learn logistic regression with the \texttt{liblinear} solver and \texttt{max\_iter=2000}.

\subsection{Variational Quantum Classifier}
\label{sec:vqc}
The VQC uses PennyLane's \texttt{default.qubit} state-vector simulator. Each PCA component is encoded with \texttt{AngleEmbedding} using $R_Y$ rotations, followed by \texttt{StronglyEntanglingLayers} of depth $L\in\{1,2,3\}$ (shallow, medium, deep), giving $3Ln$ trainable parameters for $n$ qubits (12 to 72 across the grid). Weights are initialized from $\mathcal{N}(0,0.1^2)$. The circuit outputs $\langle Z_0\rangle\in[-1,1]$, with zero as the decision threshold. Labels are mapped to $\{-1,+1\}$ and optimized with the mean squared error
\begin{equation}
\mathcal{L}(\theta)=\frac{1}{|B|}\sum_{i\in B}\bigl(y_i-\langle Z_0\rangle_{x_i,\theta}\bigr)^2,
\label{eq:loss}
\end{equation}
using Adam with learning rate 0.1, batch size 16, and 15 epochs. The loss is computed in the same forward evaluation as the gradient, so reporting it adds no second circuit evaluation to the workload. Optimization settings are fixed across the grid, so qubit count and depth are not confounded with different amounts of training.

\subsection{Runtime and Energy Accounting}
\label{sec:timing}
Wall-clock time is measured with \texttt{time.perf\_counter} around classifier fitting and validation prediction. TF-IDF and PCA are computed once per configuration \emph{outside} the timed window and shared by both arms, so the timed quantity is the classifier alone under one definition applied symmetrically. Every run is also wrapped in CodeCarbon's \texttt{OfflineEmissionsTracker} \citep{courty2024codecarbon} with \texttt{country\_iso\_code=IND}, \texttt{tracking\_mode="machine"}, and a one-second power-sampling interval; the offline tracker takes an explicit country setting instead of inferring one from IP geolocation. The timer starts before the tracker is constructed, so every timed run, classical or VQC, includes the tracker's start-up overhead. CodeCarbon values are software-based estimates, not physical power-meter readings; if tracking fails, the notebook records \texttt{NaN} rather than fabricating a value.

The main experiment comprises 59 logged rows: 39 base seed-42 rows (27 VQC, 9 unique PCA-matched classical baselines, and 3 full-TF-IDF baselines) plus 20 targeted seed-43/44 reruns (Section~\ref{sec:seeds}). Because a single classical fit lasts only about 0.06--0.10~s, its single-run CodeCarbon estimate is not a reliable basis for a cross-model energy ratio. A separate benchmark therefore repeats each of the 12 unique classical fits 1,000 times inside one tracked workload and records normalized estimated energy per fit; these estimates remain model-based.

\subsection{Experimental Grid and Logging}
\label{sec:grid}
The base grid is 3 training-set sizes $\times$ 3 qubit counts $\times$ 3 depths $=27$ VQC runs, matched to the nine unique PCA-classical configurations plus three full-TF-IDF baselines. Each run appends its full record (accuracy, F1, balanced accuracy, prediction rates, wall time, energy, CO$_2$, hardware fingerprint) to a persistent results table; validation predictions, trained weights, and per-epoch training losses are saved, so an interrupted session can resume, and mixed hardware sessions would be visible in the logged fingerprints. We flag a run as \emph{degenerate} when its positive-prediction rate is below 10\% or above 90\%; this is a descriptive diagnostic threshold, not a statistical criterion.

\subsection{Targeted Multi-Seed Check}
\label{sec:seeds}
Each base configuration runs once at seed 42. Six of the 27 matched VQC comparisons favor the VQC at that seed, with positive deltas of 0.9 to 5.7 accuracy points. These six VQC configurations, with four distinct matched classical baselines, are rerun at seeds 43 and 44, giving three seeds per selected comparison and 20 additional rows. The training subsets stay fixed; the seed changes VQC weight initialization, minibatch order, the randomized PCA solver, and classical solver state.

We report the VQC seed-to-seed standard deviation and the standard deviation of the paired VQC-minus-classical deltas. A configuration passes the \emph{descriptive} robustness rule only when its mean paired delta is positive and larger than its paired-delta seed standard deviation. This rule is not a significance test. Because the selection targets seed-42 winners, it is subject to selection effects and regression toward the mean, which we report rather than hide.

\subsection{Paired Statistical Analysis}
\label{sec:stats}
For all 27 seed-42 matched comparisons we run exact McNemar tests on the discordant prediction counts over the fixed validation set \citep{dietterich1998approximate}, with Benjamini--Hochberg correction across the 27 tests \citep{benjamini1995controlling}, and a 3,000-replicate paired bootstrap over validation examples for the accuracy difference. These tests evaluate the seed-42 models on one fixed set, whereas the multi-seed analysis (Section~\ref{sec:seeds}) evaluates stability across reruns; the two answer different questions and are reported separately.

\subsection{Environment and Reproducibility}
\label{sec:env}
All 59 rows were produced on one Kaggle CPU runtime: an Intel Xeon at 2.00~GHz (2 physical, 4 logical cores) with approximately 31.35~GB RAM, running Python 3.12.13, PennyLane 0.45.1, NumPy 2.2.6, SciPy 1.15.3, scikit-learn 1.6.1, CodeCarbon 3.3.0, and \texttt{datasets} 5.0.1. A fingerprint of versions, hardware, and configuration constants is exported with the results, and a verification script recomputes every number quoted in this paper from the exported artifacts.

\section{Results}
\label{sec:results}

\subsection{Accuracy and Computational Cost}
\label{sec:headline}
\begin{figure}[t]
\centering
\includegraphics[width=\columnwidth]{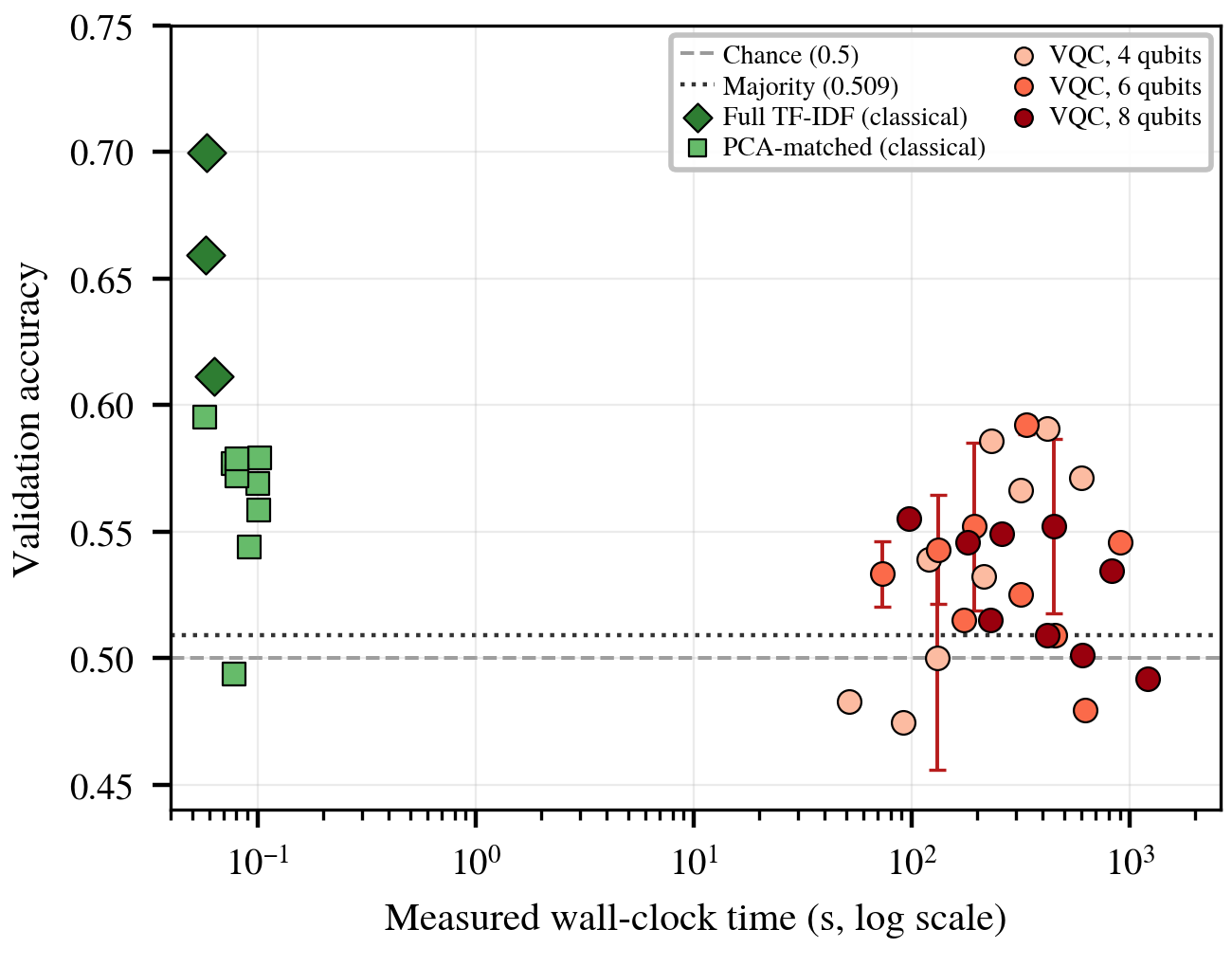}
\caption{Validation accuracy against measured wall-clock time (log scale). Configurations with additional seeds are plotted as three-seed means; VQC error bars show $\pm1$ sample standard deviation across those seeds, and the remaining points are single-seed. Horizontal lines mark random chance (0.5) and the validation majority baseline (0.5092). The VQC sits at the far right of the time axis at accuracies comparable to its matched classical baselines, while the uncompressed full-TF-IDF model is both more accurate and far cheaper.}
\label{fig:acc-vs-time}
\end{figure}

The full TF-IDF baseline improves from 61.1\% to 70.0\% as $N$ grows from 200 to 1,000 (Table~\ref{tab:headline}). Both matched arms operate under the severe PCA bottleneck and remain close to the 50.92\% majority baseline (Figure~\ref{fig:acc-vs-time}): across the 27 seed-42 VQC configurations accuracy spans 47.5\%--59.2\%, while the PCA-matched classical baselines span 49.3\%--59.5\%. At $N=500$ and $N=1000$ the VQC mean is below its matched classical mean by 5.1 and 3.1 points. The difference between the matched arms is therefore not a consistent classifier advantage. Per-configuration results for all 27 seed-42 runs are listed in Appendix~\ref{app:full}.

\begin{table}[t]
\centering
\small
\setlength{\tabcolsep}{3.2pt}
\begin{tabular}{lccccc}
\toprule
$N$ & \makecell{Full\\TF-IDF} & \makecell{PCA\\class.} & \makecell{VQC\\mean} & \makecell{$\Delta$\\(pp)} & \makecell{Median time\\ratio} \\
\midrule
200  & 0.611 & 0.525 & 0.536 & $+1.1$ & $1{,}607\times$ \\
500  & 0.659 & 0.575 & 0.524 & $-5.1$ & $3{,}158\times$ \\
1000 & 0.700 & 0.584 & 0.553 & $-3.1$ & $10{,}506\times$ \\
\bottomrule
\end{tabular}
\caption{Seed-42 accuracy and computational cost by training-set size. The VQC mean averages the nine VQC configurations (three qubit counts by three depths) at each $N$; the PCA-classical column is the corresponding mean matched baseline. $\Delta$ is VQC minus PCA-classical in accuracy points. The time ratio is VQC wall-clock time divided by its matched PCA-classical run.}
\label{tab:headline}
\end{table}

The 27 matched VQC/classical time ratios range from $886\times$ to $21{,}127\times$, with a median of $3{,}158\times$: the 27 VQC training runs total 9,663.8~s ($\approx$2.7~h), while the nine unique PCA-matched classical fits total 0.695~s under the same timing definition. Two features of that definition matter. First, each single classical run lasts only 0.057--0.102~s and so includes a non-negligible share of fixed tracker start-up overhead, which is immaterial for VQC runs of 52--1,212~s. The single-run ratios are therefore \emph{conservative}: dividing the same VQC times by the amortized per-fit times of the 1,000-repetition benchmark (1.1--1.6~ms for the PCA-matched fits) gives ratios of roughly $4.4\times10^{4}$ to $7.6\times10^{5}$, with a median of $2.1\times10^{5}$. Second, the timed window excludes TF-IDF and PCA, which both arms share. We report the single-run ratios as the headline because they apply one timing definition to both arms.

\subsection{Simulator Cost Scaling}
\label{sec:scaling}
\begin{figure*}[t]
\centering
\includegraphics[width=0.92\textwidth]{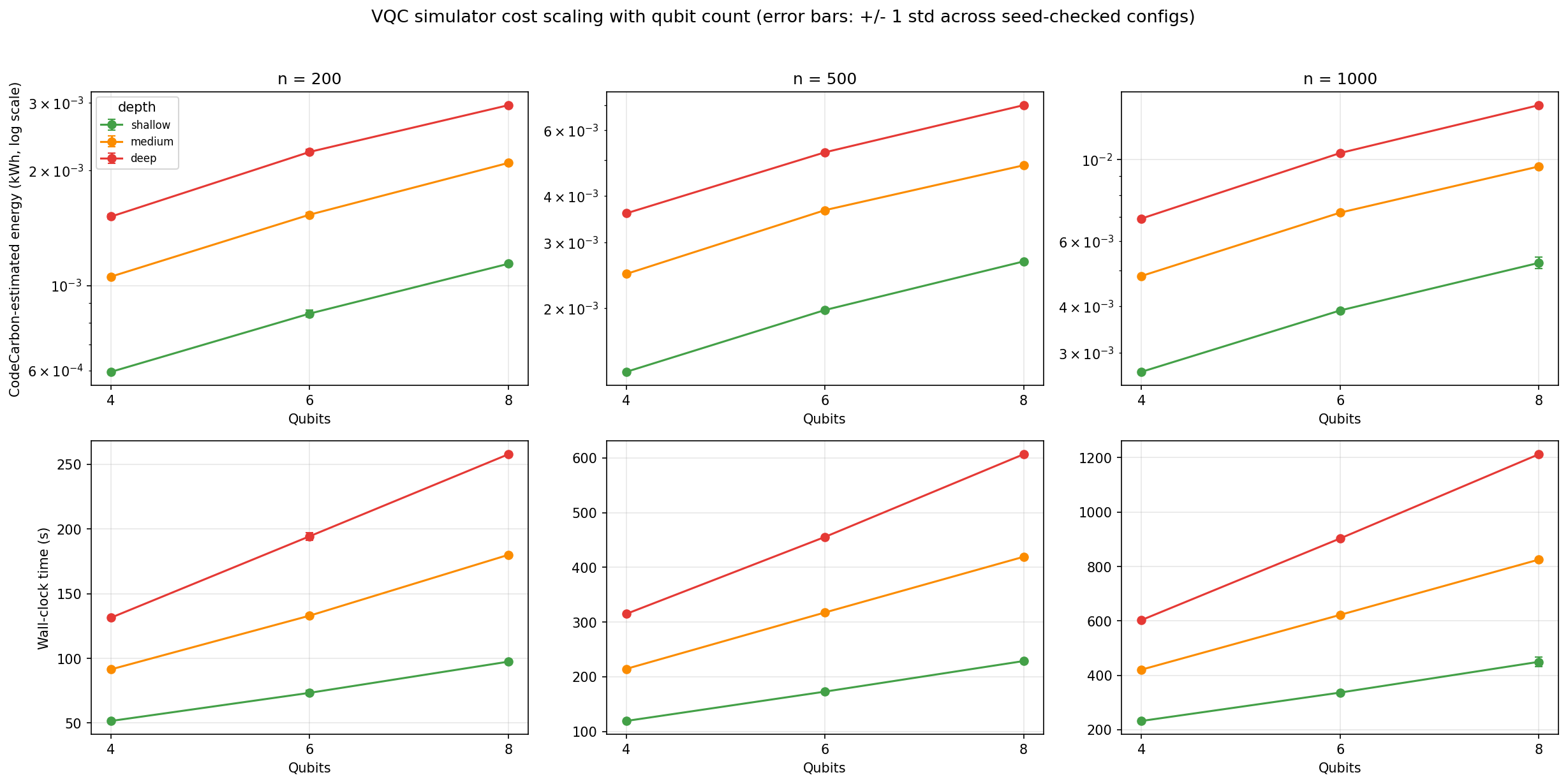}
\caption{VQC simulator cost scaling with qubit count. Columns correspond to $N=200$, 500, and 1000; colors identify circuit depth. The top row shows CodeCarbon-estimated energy and the bottom row measured wall-clock time; the rows carry nearly the same information because estimated energy is close to a constant power times elapsed time. Points are means over available seeds, and error bars (visible only for the seed-checked configurations) show $\pm1$ standard deviation. These are within-VQC measurements for the tested simulator and CPU, not hardware-independent scaling laws.}
\label{fig:scaling}
\end{figure*}

Figure~\ref{fig:scaling} and Table~\ref{tab:scaling} decompose VQC training time by qubit count, depth, and training-set size. Three regularities emerge.

\emph{Width.} At fixed $N$, going from 4 to 8 qubits multiplies mean training time by only 1.93--1.98$\times$ (4$\rightarrow$6 $\approx$ 1.5$\times$, 6$\rightarrow$8 $\approx$ 1.3$\times$), while the state vector grows from 16 to 256 amplitudes, a factor of 16.

\emph{Depth.} Deep (3-layer) circuits take 2.6--2.7$\times$ the wall time of shallow (1-layer) ones at every $N$.

\emph{Data.} With batch size 16, the number of optimizer steps is $\lceil N/16\rceil\times15$ (195, 480, and 945 for $N=200$, 500, 1000), and the mean time per step, averaged over depths, is nearly independent of $N$: 0.44--0.47~s at 4 qubits, 0.66--0.70~s at 6, and 0.87--0.91~s at 8.

Within the tested 4--8 qubit range, per-step time is well approximated by a linear function of the parameter count: across the nine (qubit, depth) cells, mean per-step time is fit by $0.053 + 0.0172\times(3Ln)$ seconds with $R^2=0.999$. This is an empirical fit to nine cells on one CPU and software stack, not a scaling law. Linear-in-gates scaling is what per-gate simulation and Python dispatch overhead would produce, and it is not what amplitude arithmetic growing as $2^{n}$ would produce at $n\leq8$; we did not profile the simulator, so we treat this mechanism as an interpretation. The exponential regime is simply not reached at this width, and these measurements say nothing about where it begins. For budgeting, total simulator cost here is approximately the number of optimizer steps times a per-step cost set by width and depth.

\begin{table}[t]
\centering
\small
\setlength{\tabcolsep}{4pt}
\begin{tabular}{rccccc}
\toprule
$N$ & \makecell{Optimizer\\steps} & \makecell{4\\qubits} & \makecell{6\\qubits} & \makecell{8\\qubits} & \makecell{$4\rightarrow8$\\factor} \\
\midrule
200  & 195 & 92   & 136  & 178  & $1.94\times$ \\
500  & 480 & 216  & 315  & 418  & $1.93\times$ \\
1000 & 945 & 419  & 620  & 827  & $1.98\times$ \\
\bottomrule
\end{tabular}
\caption{Seed-42 VQC wall-clock training time in seconds, averaged over the three circuit depths, by training-set size and qubit count. Optimizer steps are $\lceil N/16\rceil\times15$. The mean time per step, likewise averaged over depths, is nearly constant across $N$ (0.44--0.47, 0.66--0.70, and 0.87--0.91~s at 4, 6, and 8 qubits), so total cost scales approximately linearly with training-set size at fixed batch size.}
\label{tab:scaling}
\end{table}

\subsection{Accuracy Across Training-Set Size and Training Dynamics}
\label{sec:accsize}
\begin{figure*}[t]
\centering
\includegraphics[width=0.92\textwidth]{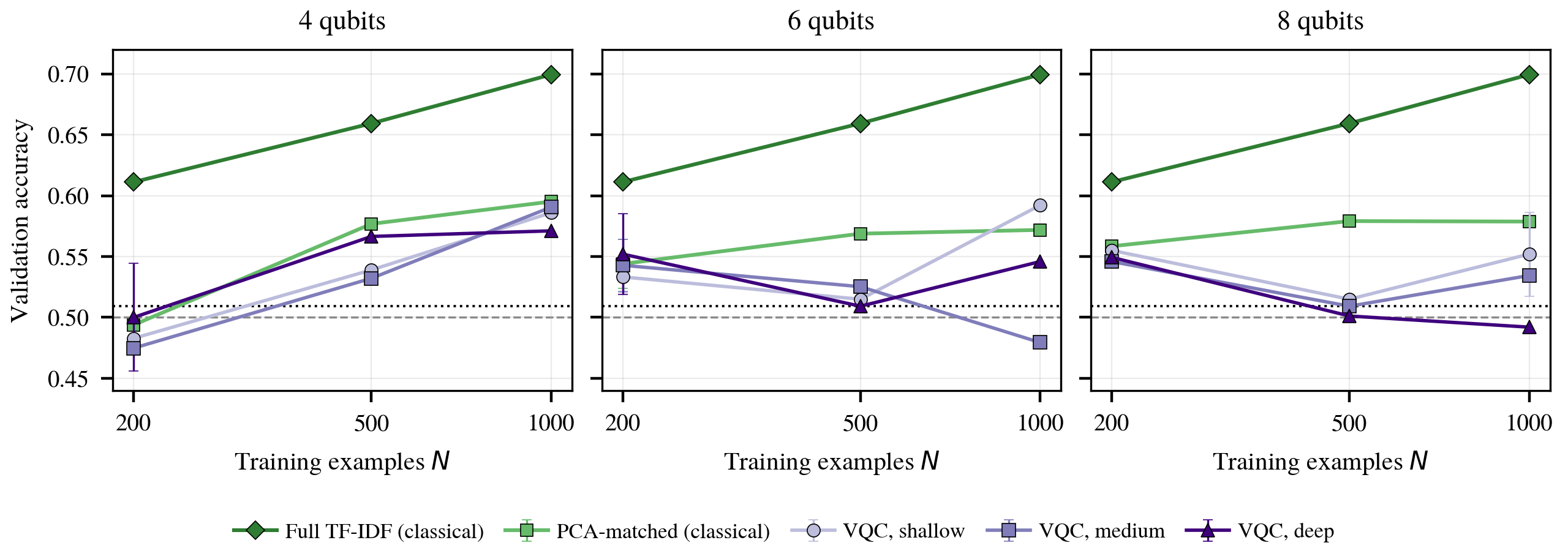}
\caption{Validation accuracy against training-set size, one panel per qubit count. Each panel shows the full-TF-IDF reference, the PCA-matched classical baseline at that qubit count, and the three VQC depths; configurations with additional seeds are three-seed means with $\pm1$ SD bars. The dashed line marks chance (0.5) and the dotted line the majority baseline (0.5092).}
\label{fig:accsize}
\end{figure*}

Figure~\ref{fig:accsize} shows accuracy against $N$ for each qubit count. The full-TF-IDF reference climbs steadily (61.1\%$\rightarrow$70.0\%), so the task and pipeline support well-above-chance accuracy when the representation is uncompressed; the PCA-matched baseline reaches 57.6\%--59.5\% at $N=1000$. The VQC improves with $N$ only at 4 qubits, where its mean accuracy over the three depths rises from 50.3\% to 54.6\% to 58.3\% and tracks the matched classical baseline (49.3\%, 57.7\%, 59.5\%). At 6 and 8 qubits the VQC means are 55.6\%, 51.6\%, 53.8\% and 55.0\%, 50.8\%, 53.9\% for $N=200$, 500, 1000: no monotonic trend, and lower at $N=1000$ than at $N=200$ by 1.8 and 1.1 points. Across the grid, 5 of the 9 (qubit, depth) cells are more accurate at $N=1000$ than at $N=200$, with a mean change of $+1.7$ points.

\begin{figure*}[t]
\centering
\includegraphics[width=0.92\textwidth]{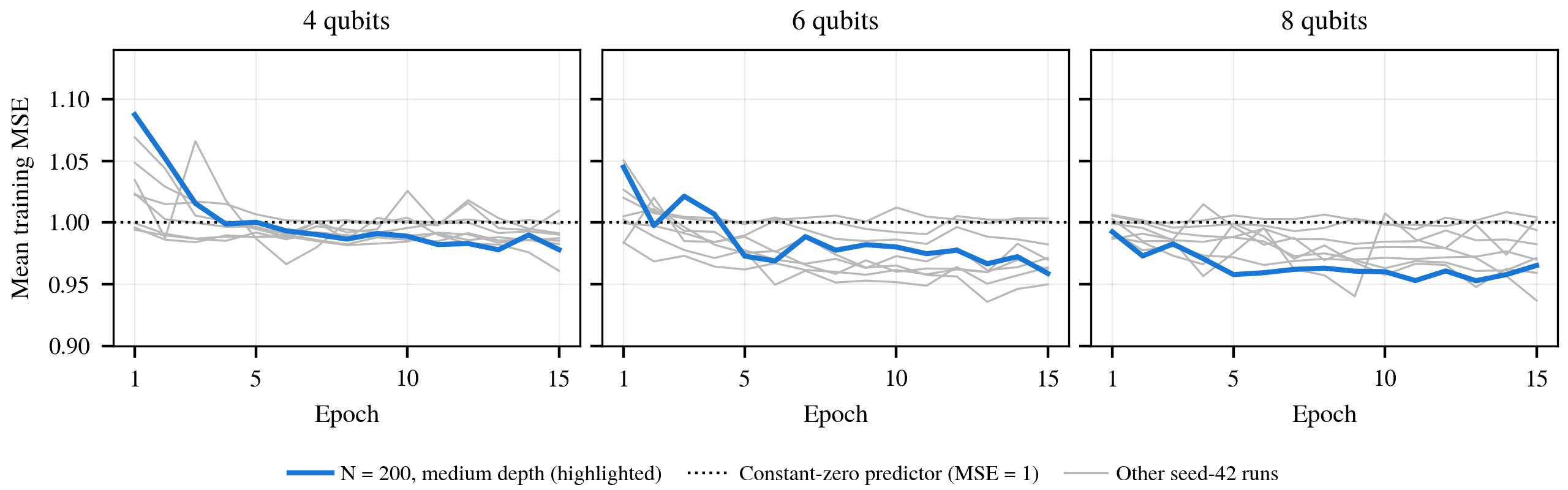}
\caption{Per-epoch mean training loss for the 27 seed-42 VQC runs, one panel per qubit count (all three training-set sizes and depths in each panel; the $N=200$, medium-depth run is highlighted). The y-axis is shared across panels. The dotted line is the loss of a constant-zero predictor on balanced $\pm1$ labels; across all 39 VQC rows the final-epoch loss lies between 0.93 and 1.01.}
\label{fig:loss}
\end{figure*}

Training loss tells the same story (Figure~\ref{fig:loss}). Across all 39 VQC rows the final-epoch mean training loss lies between 0.93 and 1.01, and it decreased from the first to the last epoch in 38 of 39 runs, by at most 0.15. With labels in $\{-1,+1\}$ and an exactly balanced training subset, a constant-zero predictor has a mean squared error of exactly 1, so the circuits fit the training labels only slightly better than a constant. We report this as an empirical property of the fixed 15-epoch, single-learning-rate protocol. Under this compressed representation, additional training examples do not reliably become additional VQC accuracy.

\subsection{Seed Robustness of Apparent Wins}
\label{sec:robust}
The targeted reruns show that single-seed improvements are unstable (Table~\ref{tab:variance}). Three of the six selected configurations have a negative three-seed mean delta; two remain positive but do not exceed their paired-delta seed variability. Only one passes the descriptive robustness rule.

\begin{table}[t]
\centering
\scriptsize
\setlength{\tabcolsep}{2.6pt}
\begin{tabular}{ccclrrc}
\toprule
$N$ & $q$ & depth & \makecell{Seed-42\\$\Delta$} & \makecell{3-seed mean\\$\Delta$} & \makecell{Paired\\$\Delta$ SD} & \makecell{Descriptive\\robust} \\
\midrule
200  & 4 & deep    & $+5.7$ & $+0.6$  & $4.5$ & no \\
200  & 6 & medium  & $+3.9$ & $-0.1$  & $3.5$ & no \\
200  & 6 & deep    & $+3.6$ & $+0.8$  & $3.4$ & no \\
200  & 6 & shallow & $+2.5$ & $-1.1$  & $3.2$ & no \\
1000 & 6 & shallow & $+1.3$ & $+2.0$  & $0.8$ & \textbf{yes} \\
1000 & 8 & shallow & $+0.9$ & $-2.7$  & $3.1$ & no \\
\bottomrule
\end{tabular}
\caption{The six configurations whose VQC accuracy exceeded its matched PCA-classical baseline at seed 42. $q$ is the qubit count (equal to the number of PCA components); shallow, medium, and deep denote 1, 2, and 3 circuit layers. Deltas are accuracy-point differences (VQC $-$ classical). The descriptive robustness rule requires a positive three-seed mean delta larger than its paired-delta seed standard deviation.}
\label{tab:variance}
\end{table}

The sole descriptive survivor, $N=1000$ with 6 qubits and shallow depth, has a three-seed mean delta of $+2.0$ points and a paired-delta standard deviation of $0.8$ points. However, its seed-42 paired bootstrap 95\% confidence interval is $[-1.7,+4.2]$ points and the exact McNemar test gives $p=0.450$ (BH-adjusted $p=0.639$). It is thus a descriptive sensitivity result, not a statistically established advantage. The converse case is instructive: the seed-42 $N=200$, 4-qubit, deep configuration is statistically favorable at that seed after BH correction ($p_{\mathrm{BH}}=0.0017$, delta $+5.7$ points) yet falls to a three-seed mean delta of $+0.6$ points once two more seeds are run. A single-seed win, however significant it looks on one fixed validation set, is not a settled effect.

\subsection{Paired Statistical Tests}
\label{sec:tests}
Across all 27 seed-42 comparisons, 11 have BH-adjusted McNemar $p<0.05$: one favors the VQC (the $N=200$, 4-qubit, deep configuration just discussed) and ten favor the PCA-matched classical baseline, with differences as large as $-9.6$ points ($N=1000$, 6 qubits, medium depth, $p_{\mathrm{BH}}=3.8\times10^{-4}$). The paired bootstrap agrees: all 11 intervals exclude zero in the same direction. Two further comparisons have bootstrap intervals that exclude zero without a BH-significant McNemar test: $N=500$, 4 qubits, shallow ($-3.8$ points, $[-7.3,-0.2]$, $p_{\mathrm{BH}}=0.107$) favors the classical baseline, and $N=200$, 6 qubits, medium ($+3.9$ points, $[+0.2,+7.7]$, $p_{\mathrm{BH}}=0.107$) favors the VQC\@. Both VQC-favoring intervals belong to the six seed-checked configurations and neither is robust across seeds (three-seed mean deltas $+0.6$ and $-0.1$ points; Table~\ref{tab:variance}). The classical-favoring differences, by contrast, were not rerun at other seeds, so they are single-seed results too. Read together, the statistical picture is that where seed-42 differences are detectable they mostly favor the classical baseline, while the VQC's apparent leads do not survive the multi-seed check.

\subsection{Prediction Behavior and Degenerate Outputs}
\label{sec:pred}
Accuracy can conceal unstable prediction behavior, so we report it explicitly. Across the 39 VQC rows the positive-prediction rate ranges from 0.6\% to 98.3\%, and three runs are degenerate under our rule (rate below 10\% or above 90\%): $N=500$, 8 qubits, deep, seed 42 (98.3\% positive); $N=1000$, 8 qubits, deep, seed 42 (0.6\% positive); and $N=200$, 4 qubits, deep, seed 44 (2.8\% positive). All three use the deepest circuit.

The highest-F1 configuration in the grid shows why F1 alone is a poor summary here. The $N=500$, 8-qubit, deep model reaches F1 $=66.6\%$, yet its accuracy is 50.1\%, its balanced accuracy 49.2\%, and it labels 98.3\% of validation sentences positive (Table~\ref{tab:confusion}). It is essentially a majority-class predictor whose F1 is inflated by the all-positive prediction. We therefore treat F1 as a diagnostic at best in this regime and report prediction balance alongside accuracy.

\begin{table}[t]
\centering
\small
\begin{tabular}{lcc}
\toprule
 & Pred.\ negative & Pred.\ positive \\
\midrule
True negative (428) & 4  & 424 \\
True positive (444) & 11 & 433 \\
\bottomrule
\end{tabular}
\caption{Confusion matrix on the 872-sentence validation set for the highest-F1 VQC configuration ($N=500$, 8 qubits, deep, seed 42). The model predicts positive for 857 of 872 sentences (98.3\%); F1 is 66.6\% while accuracy is 50.1\% and balanced accuracy 49.2\%.}
\label{tab:confusion}
\end{table}

\subsection{Energy and CO\texorpdfstring{$_2$}{2} Estimates}
\label{sec:energy}
CodeCarbon's role in this study is descriptive. Its VQC estimates accompany the wall-clock measurements and support the within-VQC scaling in Figure~\ref{fig:scaling}, but they are not physical electricity measurements, and in this environment they add little beyond runtime. The implied mean power (estimated energy divided by tracked duration) lies between 40.8 and 42.1~W for every VQC run and between 40.3 and 42.0~W for every repeated-benchmark workload, and VQC estimated energy correlates with wall-clock time at $r=0.99997$. The estimates are thus close to a constant power times elapsed time.

Single-run classical windows, by contrast, imply powers from 10 to 79~W over durations of a fraction of a second, which is why we do not build a cross-model energy ratio from them. The 1,000-repetition classical benchmark gives normalized estimates of approximately $1.27\times10^{-8}$ to $3.04\times10^{-8}$~kWh per fit under the same software model. The 27 seed-42 VQC runs sum to an estimated 0.1115~kWh and 0.0795~kg CO$_2$, which implies about 0.71~kg CO$_2$ per kWh under the India grid-intensity assumption. We report these values with their provenance rather than as headline claims.

\section{Discussion}
\label{sec:discussion}

\paragraph{A resource asymmetry, not an accuracy race.}
In the tested stack, VQC training is roughly three to four orders of magnitude slower than a classical classifier on the same reduced representation, while only 6 of the 27 single-seed accuracy differences favor the VQC (median delta $-1.8$ points). Reruns show that most apparent wins regress, and the single descriptive survivor is not statistically established. This does not say that VQCs cannot classify sentiment; it says that here the simulator's cost dwarfs that of its matched classical competitor without a reliable accuracy benefit.

\paragraph{What the low accuracies do and do not mean.}
The 4--8-dimensional PCA representation retains only 3.4--5.9\% of the variance of the 3,000-feature TF-IDF space and holds both matched arms near the majority baseline, while full TF-IDF reaches 70.0\% at $N=1000$ on the same subsets. The experiment therefore does not show that VQCs are intrinsically poor sentiment classifiers; it shows what happens when this simulator-based VQC, under a fixed protocol, is trained on this low-dimensional representation. Matching dimensionality avoids confounding classifier with representation, but it does not settle questions of quantum advantage either.

\paragraph{Scaling at small width.}
The 4$\rightarrow$8 qubit factor of about $2\times$, with roughly proportional depth and data scaling, is a calibration point for near-term QNLP practice. The $2^{n}$ state-vector cost is the right mental model for large $n$, but within the tested range per-step cost is well approximated by a linear function of the parameter count. For budgeting prototypes, simulator time can be modeled as optimizer steps times a per-step cost that grows gently with width and depth, until the exponential term takes over and the same experiment becomes infeasible rather than merely slow.

\paragraph{Reporting implications.}
The study suggests four reporting practices for simulator-based QNLP\@. First, state the timing definition, apply it to quantum and classical arms alike, and say what is excluded (here, shared TF-IDF and PCA) and what overhead is included. Second, use matched-dimensionality baselines and state the compression severity (here, explained variance). Third, verify apparent wins across seeds: our one statistically significant seed-42 win regressed to $+0.6$ points over three seeds. Fourth, when energy estimates are reported, state the tool's assumptions (grid intensity, power model) and prefer wall-clock time as the primary metric when the estimates are, as here, a near-constant multiple of it.

\section{Conclusion}
\label{sec:conclusion}
We asked what it costs to simulate a small quantum sentence classifier when the classical comparison receives the same compressed input. On one dataset, representation, ansatz, simulator, and CPU environment, the answer is 886--21,127$\times$ the wall-clock time of the matched classical fit (median 3,158$\times$; up to $7.6\times10^{5}\times$ against amortized per-fit cost) with no reliable accuracy gain: the VQC's single-seed wins largely disappear under reruns, and the statistically detectable differences mostly favor the classical baseline. Simulator cost is itself an experimental result and should be reported alongside accuracy in QNLP studies, particularly when a quantum model is compared with a classical baseline on a matched representation.

\section*{Limitations}

\paragraph{Estimated energy.}
CodeCarbon's offline tracker reports software-based energy and CO$_2$ estimates from the observed runtime and its power model. The \texttt{country\_iso\_code=IND} setting is an explicit grid-intensity assumption, not a measurement of the host's electricity source. Because the VQC and benchmark estimates imply a nearly constant power, they carry essentially the same information as wall-clock time.

\paragraph{Measurement scale.}
The classical single-fit workloads are extremely short and their timed windows include tracker start-up, so we do not interpret their raw single-run CodeCarbon values as precise energy observations, and the single-run time ratios are conservative. The repeated benchmark improves the measurement window but remains model-based.

\paragraph{Single environment.}
All 59 rows come from one Kaggle CPU runtime (Section~\ref{sec:env}). Results describe this software and hardware environment, not a hardware-independent simulator law; GPU-accelerated or JIT-compiled simulators, different thread counts, or different power models could shift both absolute times and scaling factors. We did not profile the simulator, so the overhead interpretation of the scaling results is untested.

\paragraph{Partial seed coverage.}
Only the six seed-42 VQC winners were rerun at seeds 43 and 44. The other 21 VQC configurations, and all classical-favoring differences, rest on single-seed estimates. The seed check samples optimization and solver randomness within fixed training subsets; it does not test sensitivity to a different draw of SST-2 training examples, and it does not estimate the win rate over all 27 configurations.

\paragraph{Representation and model scope.}
We test one TF-IDF/PCA representation, one ansatz (\texttt{StronglyEntanglingLayers}), one angle encoding, one loss and optimizer configuration, and logistic regression as the classical classifier. Other representations, classical models, simulators, or compositional QNLP models could give different accuracy and cost profiles; nothing here generalizes to grammar-aware architectures.

\paragraph{Selection.}
The highest-F1 configuration is selected post hoc and is presented only as a prediction-behavior diagnostic. The six multi-seed configurations are selected after observing seed-42 outcomes, so their rerun statistics are not an unbiased estimate over all 27 configurations.

\section*{Ethics Statement}
This study uses only a public, anonymized benchmark (SST-2) and involves no human subjects or personal data. Its energy and CO$_2$ figures are software estimates under a stated grid-intensity assumption, not physical measurements, and are presented to promote transparent cost reporting rather than to certify environmental impact. The study is a feasibility measurement on one narrow stack and should not be cited as evidence for or against quantum computing at large.

\section*{Code and Data Availability}
This arXiv submission includes selected machine-readable result tables, statistical summaries, and figure data for the reported results as ancillary files. The full reproducibility package is deposited in a versioned \href{https://doi.org/10.5281/zenodo.23117261}{Zenodo archive}. It contains the Kaggle notebook, analysis scripts, run metadata, per-run validation predictions, trained circuit weights, and other supporting files. The Zenodo record is public. Access to the deposited files is restricted, and the authors can grant it on request.

\begingroup
\raggedright
\bibliography{custom}
\endgroup

\clearpage
\onecolumn
\appendix

\section{Full Seed-42 Results}
\label{app:full}
Table~\ref{tab:full} lists every seed-42 VQC configuration with its matched PCA-classical baseline, so that each claim in Section~\ref{sec:results} can be traced to a row. All values are produced from the exported experiment log and paired-test tables by \texttt{scripts/make\_appendix\_table.py}.

\begin{table}[!ht]
\centering
\small
\setlength{\tabcolsep}{6pt}
\begin{tabular}{rcllrrrrrr}
\toprule
$N$ & $q$ & depth & \makecell{VQC\\acc.\ (\%)} & \makecell{PCA-class.\\acc.\ (\%)} & \makecell{$\Delta$\\(pp)} & \makecell{McNemar\\$p_{\mathrm{BH}}$} & \makecell{VQC\\time (s)} & \makecell{Time\\ratio ($\times$)} & \makecell{VQC pos.\\rate (\%)} \\
\midrule
200 & 4 & shallow & 48.3 & 49.3 & $-1.0$ & 0.735 & 51.6 & 907 & 55.2 \\
200 & 4 & medium & 47.5 & 49.3 & $-1.8$ & 0.545 & 91.5 & 1{,}607 & 32.6 \\
200 & 4 & deep$^\dagger$ & 55.0 & 49.3 & $+5.7$ & 0.0017 & 132.9 & 2{,}334 & 27.3 \\
200 & 6 & shallow$^\dagger$ & 54.8 & 52.3 & $+2.5$ & 0.337 & 75.6 & 886 & 53.9 \\
200 & 6 & medium$^\dagger$ & 56.2 & 52.3 & $+3.9$ & 0.107 & 134.1 & 1{,}573 & 60.3 \\
200 & 6 & deep$^\dagger$ & 55.8 & 52.3 & $+3.6$ & 0.159 & 196.9 & 2{,}310 & 54.0 \\
200 & 8 & shallow & 55.5 & 55.8 & $-0.3$ & 0.912 & 97.4 & 966 & 71.3 \\
200 & 8 & medium & 54.6 & 55.8 & $-1.3$ & 0.664 & 180.0 & 1{,}783 & 48.4 \\
200 & 8 & deep & 54.9 & 55.8 & $-0.9$ & 0.735 & 257.7 & 2{,}554 & 35.0 \\
500 & 4 & shallow & 53.9 & 57.7 & $-3.8$ & 0.107 & 119.3 & 1{,}548 & 42.9 \\
500 & 4 & medium & 53.2 & 57.7 & $-4.5$ & 0.013 & 214.4 & 2{,}783 & 51.1 \\
500 & 4 & deep & 56.7 & 57.7 & $-1.0$ & 0.664 & 314.9 & 4{,}087 & 50.7 \\
500 & 6 & shallow & 51.5 & 56.9 & $-5.4$ & 0.017 & 172.9 & 1{,}720 & 29.5 \\
500 & 6 & medium & 52.5 & 56.9 & $-4.4$ & 0.039 & 317.5 & 3{,}158 & 37.4 \\
500 & 6 & deep & 50.9 & 56.9 & $-6.0$ & 0.0087 & 455.5 & 4{,}531 & 18.8 \\
500 & 8 & shallow & 51.5 & 57.9 & $-6.4$ & 0.0070 & 228.8 & 2{,}244 & 26.9 \\
500 & 8 & medium & 50.9 & 57.9 & $-7.0$ & 0.0042 & 419.2 & 4{,}112 & 21.1 \\
500 & 8 & deep & 50.1 & 57.9 & $-7.8$ & 0.0054 & 606.5 & 5{,}949 & \textbf{98.3} \\
1{,}000 & 4 & shallow & 58.6 & 59.5 & $-0.9$ & 0.664 & 232.1 & 4{,}044 & 35.9 \\
1{,}000 & 4 & medium & 59.1 & 59.5 & $-0.5$ & 0.838 & 420.6 & 7{,}330 & 32.7 \\
1{,}000 & 4 & deep & 57.1 & 59.5 & $-2.4$ & 0.262 & 602.9 & 10{,}506 & 68.1 \\
1{,}000 & 6 & shallow$^\dagger$ & 58.8 & 57.6 & $+1.3$ & 0.639 & 336.5 & 5{,}879 & 30.4 \\
1{,}000 & 6 & medium & 47.9 & 57.6 & $-9.6$ & $3.8\times10^{-4}$ & 622.0 & 10{,}867 & 45.9 \\
1{,}000 & 6 & deep & 54.6 & 57.6 & $-3.0$ & 0.352 & 902.9 & 15{,}775 & 45.9 \\
1{,}000 & 8 & shallow$^\dagger$ & 59.2 & 58.3 & $+0.9$ & 0.735 & 442.6 & 7{,}713 & 30.3 \\
1{,}000 & 8 & medium & 53.4 & 58.3 & $-4.8$ & 0.037 & 825.2 & 14{,}380 & 17.0 \\
1{,}000 & 8 & deep & 49.2 & 58.3 & $-9.1$ & 0.0014 & 1{,}212.4 & 21{,}127 & \textbf{0.6} \\
\bottomrule
\end{tabular}
\caption{All 27 seed-42 VQC configurations. $q$ is the qubit count. $\Delta$ is VQC minus the matched PCA-classical accuracy in points. $p_{\mathrm{BH}}$ is the Benjamini--Hochberg-adjusted exact McNemar $p$-value across the 27 comparisons. The time ratio is VQC wall-clock time divided by the matched classical run (single-run definition of Section~\ref{sec:timing}). Bold positive-prediction rates mark degenerate runs (below 10\% or above 90\%); a third degenerate run occurs at seed 44 ($N=200$, 4 qubits, deep; 2.8\%). $^\dagger$ marks the six configurations rerun at seeds 43 and 44.}
\label{tab:full}
\end{table}

\end{document}